\documentclass[prb,showpacs,preprintnumbers,twocolumn,amsmath,amssymb,superscriptaddress]{revtex4-1}

\usepackage[T1]{fontenc}
\usepackage[applemac]{inputenc}
\usepackage[english]{babel}
\usepackage{graphics}
\usepackage{amssymb}
\usepackage{amsmath}
\usepackage{overpic}
\usepackage{color}
\usepackage{epstopdf}

\newcommand{\be}{\begin{equation}}
\newcommand{\ee}{\end{equation}}
\newcommand{\bea}{\begin{eqnarray}}
\newcommand{\eea}{\end{eqnarray}}

\def\a{\alpha}

\def\d{\delta}

\def\m{\mu}
\def\l{\lambda}

\def\s{\sigma}

\def\O{\Omega}

\def\ra{\rightarrow}

\def\pd{\partial}
\def\nb{\nabla}

\def\bq{{\bf q}}
\def\br{{\bf r}}

\def\bA{{\bf A}}

\def\bx{{\bf x}}

\def\nn{\nonumber}
\def\lb{\label}
\def\pref#1{(\ref{#1})}

\newcount\bozza \bozza=1
\ifnum\bozza=0
\newdimen\shift \shift=-2truecm
\def\lb#1{%
{\label{#1}\rlap{\kern\shift{$\scriptstyle#1$}}}}
\else\def\lb#1{\label{#1}} \fi

\begin{document}

\title{Optical signatures of the superconducting Goldstone mode in granular aluminum: experiments and theory}
\author{Uwe S. Pracht}
\affiliation{1.Physikalisches Institut, Universit\"at Stuttgart, Pfaffenwaldring 57, 70569 Stuttgart, Germany}
\author{Tommaso Cea}
\affiliation{Graphene Labs, Fondazione Istituto Italiano di Tecnologia, Via Morego, 16163 Genova, Italy}
\affiliation
{ISC-CNR and Department of Physics, Sapienza University of Rome, P.le Aldo Moro 5, 00185, Rome, Italy}
\author{Nimrod Bachar}
\affiliation{DPMC, University of Geneva, 24 Quai Ernest-Ansermet
CH-1211, Geneva, Switzerland}
\author{Guy Deutscher}
\affiliation{Raymond and Beverly Sackler School of Physics and Astronomy, Tel Aviv University, Israel}
\author{Eli Farber}
\affiliation{Dep. of Physics and Dep. of Electrical and Electronic Engineering,  Ariel University, P.O.B.  3, Ariel 40700, Israel}
\author{Martin Dressel}
\affiliation{1.Physikalisches Institut, Universit\"at Stuttgart, Pfaffenwaldring 57, 70569 Stuttgart, Germany}
\author{Marc Scheffler}
\affiliation{1.Physikalisches Institut, Universit\"at Stuttgart, Pfaffenwaldring 57, 70569 Stuttgart, Germany}
\author{Claudio Castellani}
\affiliation
{ISC-CNR and Department of Physics, Sapienza University of Rome, P.le Aldo Moro 5, 00185, Rome, Italy}
\author{Antonio M. Garc\'ia-Garc\'ia}
\affiliation{Cavendish Laboratory, University of Cambridge, JJ Thomson Av., Cambridge, CB3 0HE, UK}
\author{Lara Benfatto}
\affiliation
{ISC-CNR and Department of Physics, Sapienza University of Rome, P.le Aldo Moro 5, 00185, Rome, Italy}


\begin{abstract}
Recent advances in the experimental growth and control of disordered thin films, heterostructures, and interfaces provide a fertile ground for 
the observation and characterisation of the collective superconducting excitations emerging below $T_c$ after breaking the $U(1)$ gauge symmetry. Here we combine THz experiments in 
a nano-structured granular Al thin film and theoretical calculations to demonstrate the existence of optically-active phase modes, which represent the Goldstone excitations of the broken gauge symmetry. By measuring the complex transmission trough the sample we identify a sizeable and temperature-dependent optical sub-gap absorption, which cannot be ascribed to quasiparticle excitations. A quantitative modelling of this material as a disordered Josephson array of nano-grains allows us to determine, with no free parameters, the structure of the spatial inhomogeneities induced by shell effects. Besides being responsible for the enhancement of the critical temperature with respect to bulk Al, already observed in the past, this spatial inhomogeneity provides a mechanism for the optical visibility of the Goldstone mode. By computing explicitly the optical spectrum of the superconducting phase fluctuations we obtain a good quantitative description of the experimental data. Our results demonstrate that nanograins arrays are a promising setting to study and control the collective superconducting excitations via optical means.

\end{abstract}

\maketitle
\section{Introduction}
Zero resistance at finite temperature in systems where momentum is not conserved, one of the defining features of a superconductor, is strictly related \cite{schrieffer,nagaosa} to the phase rigidity of the complex order parameter. Phase rigidity is a typical consequence of a spontaneous symmetry breaking: the Hamiltonian that describes the superconductor is invariant under a phase rotation of the electronic degrees of freedom but in the ground state the macroscopic order parameter chooses a particular value of the electronic phase, and therefore breaks spontaneously this rotational $U(1)$ symmetry. According to the Goldstone theorem, the  collective excitation connecting the  possible degenerate ground states must be massless at long wavelength\cite{schrieffer,nagaosa}.

In principle, this collective mode, usually termed the Goldstone mode \cite{nambu,goldstone}, should manifest as a low-energy excitation.
In the context of superconductivity it should appear as a sub-gap excitation. However, despite intensive research, it has not yet conclusively been observed experimentally. 
The standard explanation is based on the idea of Anderson\cite{Anderson1958}, proposed shortly after the BCS theory: Coulomb interactions, present in any material, boost the typical frequency of the Goldstone mode to the plasma energy scale, well above the energy gap, so that it cannot be thermally excited at temperatures below the superconducting (SC) critical temperature $T_c$. A second issue is that typical spectroscopic measurements probe the system in the long-wavelength regime where the phase mode is decoupled from the transverse electromagnetic field, so that one cannot observe it in the ac conductivity. However, these conclusions only hold for a spatially homogeneous SC state, since disorder and inhomogeneity can affect both the spectrum of the phase mode and its optical visibility. 

In recent years, due to experimental advances in the growth and control of SC thin films, the electromagnetic response of conventional superconductors have been studied with unprecedented precision and in a broad range of disorder strengths. The improved experimental resolution revealed that in some conventional $s$-wave superconductors, as NbN, InO$_x$ and granular Al, a finite absorption can be found even below the threshold $2\Delta$ for the Cooper-pair breaking, with $\Delta$ being the SC gap
\cite{armitage_prb07,driessen12,driessen13,bachar_jltp14,frydman_natphys15,samuely15,armitage_prb16,scheffler16,pracht_prb16}.
One possible interpretation\cite{stroud_prb00,cea_prb14,trivedi_prx14,seibold_cm17} of these experiments points out to the relevance of the Goldstone mode, made optically active by the spontaneous inhomogeneity of the SC ground state that has been observed in strongly-disordered films\cite{sacepe08,sacepe10,mondal10,sacepe11,chand12,kam13,roditchev13,roditchev_natphys14,samuely16,leridon_prb16,brun_review17}. This interpretation is still under debate for two main reasons. From one side, the explicit calculation of the phase-mode absorption has been performed so far\cite{stroud_prb00,cea_prb14,trivedi_prx14,seibold_cm17} under the assumption that long-range Coulomb forces should not be included. This relies on the expectation that the optical conductivity is the response to the local field, so it is irreducible with respect to the Coulomb interaction\cite{belitz_prb89,cea_prb14}, i.e. it only depends on the undressed sound-like phase spectrum. Nonetheless, disorder could still mix the reducible and irreducible response, and an explicit estimate of this effect is still lacking.  From the other side,  other mechanisms could potentially 
induce sub-gap excitations in the proximity of the  superconductor-to-insulator transition (SIT).
One interesting proposal\cite{frydman_natphys15} is that large enough disorder can push the SC system towards an effective strong-coupling regime, where the energy scale of the Higgs mode would lie below the one for Cooper-pair breaking, making it visible as a sub-gap excitation. However, explicit calculations\cite{cea_prl15} within disordered fermionic models did not find yet evidence of a sharp, subgap Higgs mode in the strong-coupling regime emerging near the SIT.  

It is then clear that  an 
experimental confirmation of the role of Goldstone modes would require a weakly-coupled and inhomogeneous superconductor, where the  inhomogeneity is not driven by too large disorder, so that the sub-gap features cannot be attributed to other unrelated effects. It would be also desirable to establish a quantitative relation between the experimental source of inhomogeneity and the parameters of the theoretical model, that presently is not known for inhomogeneities induced by impurities. A promising settings for this analysis is provided by films made of well-coupled nanograins. Indeed, recent research \cite{Bose2010,Garcia-Garcia2008,Garcia-Garcia2011,Brihuega2011,Shanenko2006,Shanenko2007} in single isolated superconducting nano-grains of Sn has found that shell effects, 
induced by fluctuations in the density of states, lead to large changes in the SC gap by very small changes in the grain size, provided that the grain shape is sufficiently symmetric.  
As a result, a Josephson array of SC nano-grains can be highly inhomogeneous even when the SC transition occurs very far from the SIT. Unlike inhomogeneities induced by impurities it is possible to know experimentally \cite{Lerer2014,Deutscher1973,Deutscher1973a,pracht_prb16}, with relative accuracy, the grain size, its distribution, and the density of grains, that control the resistivity of the material and the inhomogeneity of its SC properties. An additional advantage of the nanograins, as compared to strongly-disordered superconductors and arrays made of artificial Josephson junctions, is that in this case the capacitive charging energy of each grain dominates over the junction capacitance. In this situation the long-range part of the Coulomb interaction is automatically screened, and the phase mode remains an acoustical mode at long wavelength\cite{fazio_review01,halperin_prb89,ribeiro2014,garcia_prb14}.

A prototype system belonging to this category is granular Al. 
The typical grain size, a few nanometers, and its shape, spherical, is optimal\cite{garcia_prb14} to observe strong shell effects and consequently strong inhomogeneities in the array even in the low- resistivity region, where the average coupling among grains is strong enough to give a dc conductivity consistent with metallic behavior. An indirect proof of the emerging inhomogeneity is provided by the  experimentally observed enhancement of the critical temperature \cite{Abeles1966,Deutscher1973,Lerer2014,pracht_prb16} with respect to bulk Al. Indeed, it has been argued\cite{garcia_prb14}  that the local enhancement of the superconducting gap due to shell effects can increase, within a percolative scheme, the critical temperature of the array.

A natural question to ask is whether inhomogeneities induced by shell effects can also lead to the observation of collective Goldstone excitations. Here we answer this question positively.
More specifically, we first provide direct experimental evidence of a broad sub-gap resonance in the ac conductivity of granular Al film deep in the metallic region. In this sample corrections to the bulk mean-field due to shell effects are still substantial to induce inhomogeneity in the system, but the film is metallic enough to suppress the corrections to the critical temperature due to quantum phase fluctuations\cite{garcia_prb14}. We model granular Al as an array of nano-grains where the inhomogeneity of the Josephson coupling between grains is computed microscopically starting from the  grain-size distribution. The obtained distribution of the local coupling is then used as an input to compute both the $T_c$ of the array and the optical response of the disordered phase modes. The same model can then explain two striking experimental observations in the system: (i) the enhancement of  $T_c$  with respect to bulk Al and (ii) the emergence of an optically-active Goldstone mode, visible as an extra sub-gap absorption. The good quantitative agreement between the theoretical calculations and the experiments provides a strong support to the relevance of Goldstone modes in inhomogeneous superconductors, opening interesting perspectives for applications in artificially-designed inhomogeneous systems. 

The paper is organized as follows: In Sec. II we describe the experimental setup and the measurements in a granular Al sample in the metallic regime. Sec. III describes the theoretical model used to compute the SC properties of the arrays. In Sec. IV we compare explicitly the theoretical results with the experimental data for the THz conductivity. The concluding remarks are presented in Sec. V. Additional technical details on the theoretical calculations are provided in Appendix A and B.

\section{Experiments}

\begin{figure}[htb]
\centering
\includegraphics[width=\columnwidth]{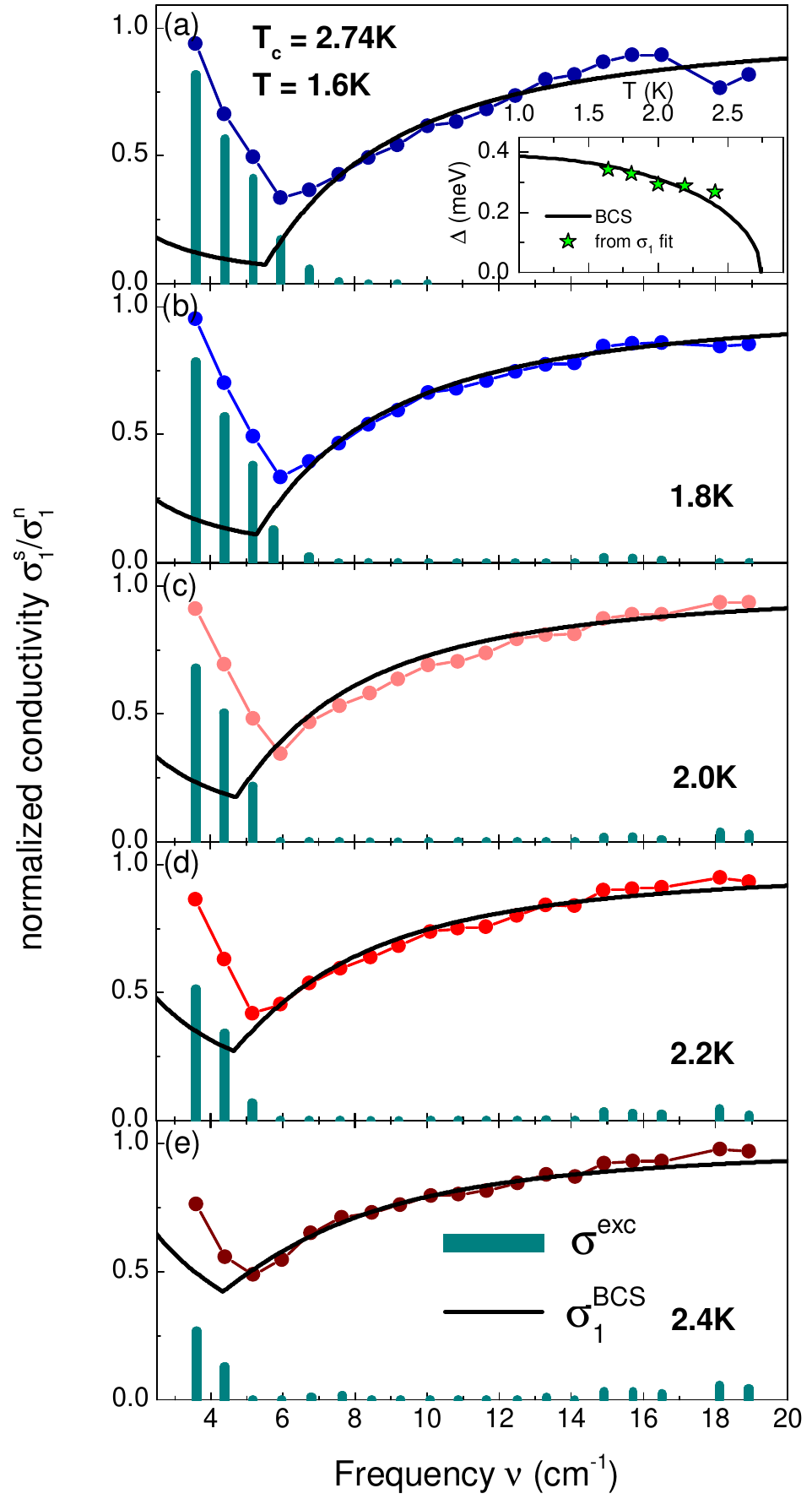}
\caption{\label{sig1_exp} Real part (full circles) of the experimentally-determined (normalized) dynamical conductivity $\sigma_1^{exp}$ versus frequency $\nu$ at various temperatures below $T_c=2.74$\,K. The solid lines are fits to the Mattis-Bardeen BCS prediction restricted to frequencies $\nu\geq 2\Delta/(hc)$. The bars document the excessive conductivity $\sigma^{exc}$, as defined in Eq.\ \pref{sexcess}. The inset of the top panel shows the temperature dependence of the SC gap $\Delta(T)$ (stars) extracted from the Mattis-Bardeen analysis of the optical conductivity, along with a BCS fit (solid line). }
\end{figure}

To access the frequency-dependent ($\nu$) complex conductivity $\sigma(\nu)=\sigma_1(\nu)+i\sigma_2(\nu)$, we performed phase-sensitive measurements \cite{Pracht2013} of the complex transmission coefficient of tunable and coherent THz radiation passing through a granular Al film of $d=40$\,nm thickness. The film was grown on a MgO$_2$ dielectric $10\times 10$\,mm$^2$  substrate held at 77\,K via thermal evaporation in partial pressure of O$_2$. The O$_2$ concentration while deposition was adjusted such to produce a granular morphology characterized by a dc-resistivity of $\rho_{dc}=0.26\times 10^{-3} \O$cm at 5\,K and a strongly enhanced $T_c$ of 2.74\,K. We simultaneously fit the amplitude and phase of the transmitted THz radiation with Fresnel equations via $\sigma_1$ and $\sigma_2$ (detailed information on the analysis procedure is found in \cite{Pracht2013}). Note that this approach is not based on any microscopic model for the charge carrier dynamics. 
The low-temperature data for several samples at different disorder levels have been shown in a previous publication \cite{pracht_prb16}.

As $\sigma_1$ equally probes the coupling of light to single-particle excitations as well as to optically-active collective modes of the order parameter, it can be difficult to clearly disentangle different absorption channels. To outline the effect of the Goldstone mode, we focus on the real (dissipative) part $\sigma_1$ and define the excess conductivity with respect to the Mattis-Bardeen BCS prediction
\be
\lb{sexcess}
\frac{\sigma^{exc}}{\sigma_{dc}}\equiv \frac{\sigma_1^{exp}-\sigma_1^{BCS}}{\sigma_{dc}}
\ee
where $\sigma_{dc}$ is the dc conductivity measured right above $T_c$. We infer $\sigma_1^{BCS}$ by fitting $\sigma_1^{exp}$ at frequencies above the energy gap, i.e. the excitations into the quasiparticle continuum, to the standard Mattis-Bardeen functional \cite{mb,pracht_prb16}, see Fig.\, \ref{sig1_exp}. Interestingly, this yields a surprisingly good description of $\sigma_1^{exp}(\nu)$ for $\nu\geq 2\Delta/(hc)$ and, at the same time, reveals (and quantifies) the notable excessive absorption $\sigma_1^{exc}(\nu)$ at frequencies below twice the SC gap. As one can see in Fig.\ \ref{sig1_exp}, as the temperature rises from $T=1.6$\,K towards $T_c=2.74$\,K, the extra absorption gets continuously suppressed, but nonetheless it still represents a significant contribution to $\sigma_1^{exp}$ aside the thermally-excited quasiparticles.
On the other hand, the remaining features of the spectra are rather conventional. For example, the inset of Fig.\ \ref{sig1_exp} shows the temperature dependence of the gap, as extracted from the Mattis-Bardeen fit.  It can be very well fitted to a BCS-ilke behavior, giving an extrapolated value at $T=0$ equal to $\Delta=0.4$ meV, so that $\Delta/T_c=1.78$ has the conventional weak-coupling value, despite the enhancement of the $T_c$ as compared to bulk Al. 

As observed before\cite{pracht_prb16} the presence of extra sub-gap absorption is a general feature of granular Al samples regardless the level of inter-grains coupling, measured by the value of the normal-state resistivity. This has to be contrasted with homogeneously disordered films of conventional superconductor\cite{frydman_natphys15,driessen12,armitage_prb16,practh2012}, where significant deviations from the Mattis-Bardeen prediction only occurs at relatively large disorder levels, where the dc conductivity displays an insulating behavior. To make a quantitative comparison let us consider for example the NbN films of Ref.\ \cite{armitage_prb16}. Here a consistent sub-gap absorption, comparable to the one reported in Fig.\ \ref{sig1_exp}, appears only for film resistance $\rho_{dc}=1.2\times 10^{-3} \O$cm, with film thickness $60-120$ nm. For our granular Al sample the sub-gap absorption can be clearly distinguished already for a resistivity value about one order of magnitude smaller, $\rho_{dc}=0.26\times 10^{-3} \O$cm (and $d=40$ nm).

\section{Theoretical model}

\subsection{Optical response of disordered phase modes}
\begin{figure}
	\centering
	\includegraphics[scale=0.1]{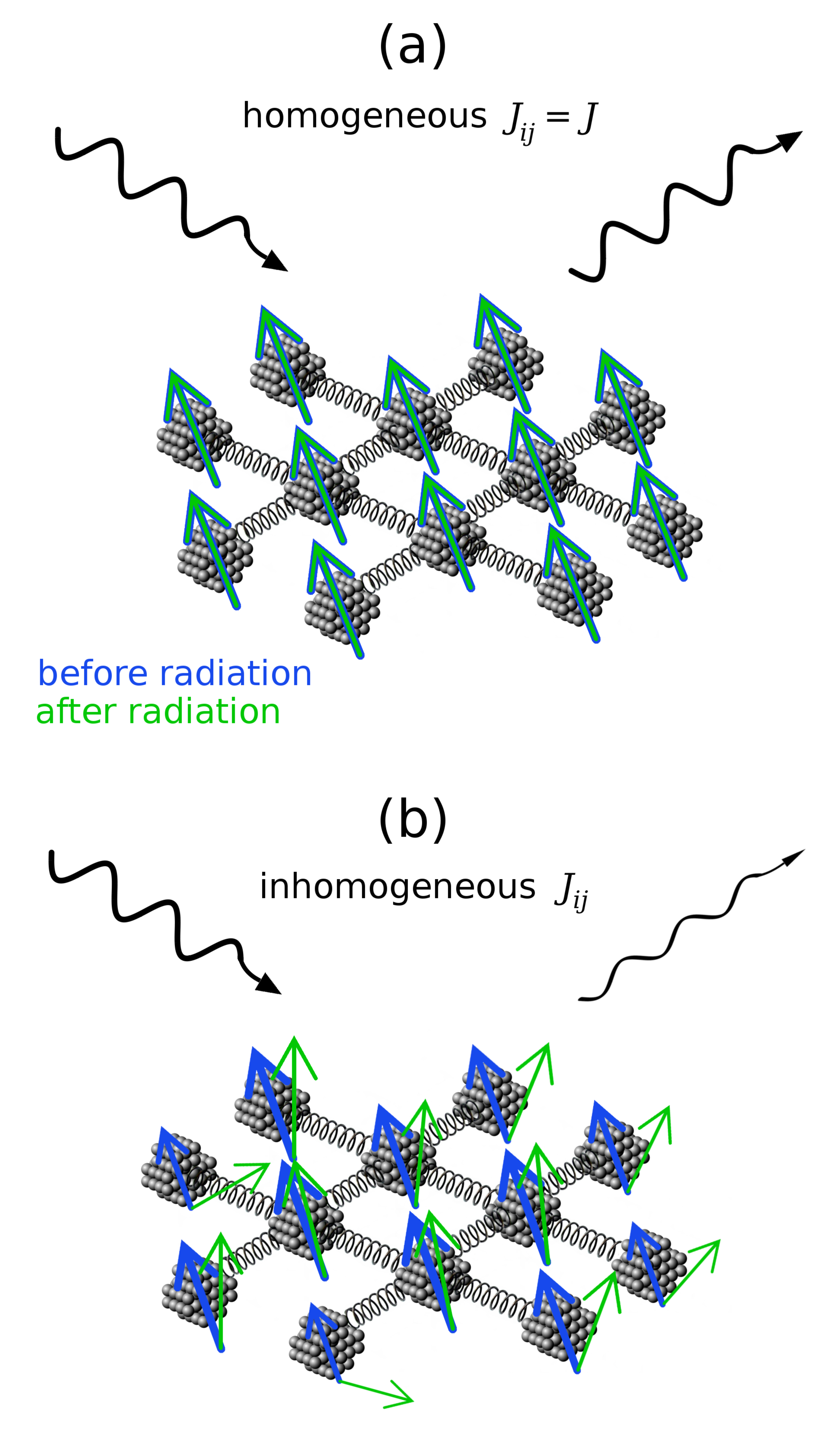}
	\caption{Sketch of the optical response of an array of Josephson junction, modeled as in Eq.\ \pref{hxy}. Here the arrows represent the local spins on the array sites, connected by springs representing the local stiffnesses $J_{ij}$. To visualise its inhomogeneity, we set the arrow length proportional to the strength of the local stiffness. In the clean case, panel (a), the phase modes are decoupled from the transverse electromagnetic field, so the spins preserve their orientation (i.e. the phase of the SC order parameter is unchanged) and the radiation is not absorbed.  On the other hand in the disordered case, panel (b), the spins respond to the incoming radiation with a local change of their relative direction that is larger when the system has lower phase rigidity (i.e. lower local $J_{ij}$). This leads to an inelastic response which absorbs part of the incoming radiation. }
	\label{fig-phasemode}
	\vspace{1cm}
\end{figure}

As a starting point for the theoretical analysis of the optical conductivity we need a proper model for the phase degrees of freedom in a inhomogeneous superconductor. Our main aim is to study low-energy properties for which it is not necessary to consider amplitude fluctuations. We will also consider the quasi-two dimensional limit for computational simplicity. A reasonable description for the phase fluctuations in a granular system  is then  provided by the two-dimensional (2D) quantum XY model \cite{matsubara,anderson59} for $S=1/2$ pseudospins with bonds on a square lattice:\\
\begin{equation}
\lb{hxy}
	\mathcal{H}_{XY}=-2\sum_{\langle i,j	\rangle}J_{ij}\left(S^+_iS^-_j+h.c.		\right)\quad,
\end{equation}
where the sum is extended to all the pairs of nearest neighbour sites and $J_{ij}$ is the hopping that depends on the 
resistance of the model and the value of the SC gap, i.e. the amplitude of order parameter,  in neighboring grains. The local gap can be  computed in the mean-field limit as a function of the grain size\cite{garcia_prb14}. We first discuss the outcomes of the model \pref{hxy} for a generic $J_{ij}$. In the second part of the section we describe the details of the calculation of $J_{ij}$. 

The  idea underlying the present approach is the well-known Anderson mapping\cite{anderson59} between a $s$-wave superconductor and the pseudospin Hamiltonian \pref{hxy}. Here the $S_i$ are spin $1/2$ operators,  such that the in-plane spin component describes the pairing operator, $S^-_i=c_{i\downarrow}c_{i\uparrow}$, while the $z$ component describes the local density, $S^z_i=\frac{1}{2}\left(\sum_\sigma c^\dagger_{i\sigma}c_{i\sigma}-1\right)$, with $c_{i\sigma}$ fermionic annihilation operators. Thus within such a spin-like picture of superconductivity the SC order appears as the spontaneous magnetization of  the $XY$ component, and $J_{ij}>0$ is the hopping amplitude, which represents the energetic gain to move a Cooper pair from a given site $j$ to a nearest neighbor site $i$.


%
%

An exact treatment of Eq.\ \pref{hxy} is beyond current analytical techniques. In order to proceed we compute the mean-field ground state, and fluctuations with respect to the mean-field ground state, by mapping the spins into bosonic operators by means of the usual
  Holstein-Primakov approximation (see Appendix B). After the mapping, the evaluation of the superconducting phase-mode spectrum is equivalent to the evaluation of the spin-wave spectrum in an ordinary ferromagnetic spin model. In practice, this is equivalent to study the following quantum phase-only model:
\be
\lb{quant}
H_{HP}\simeq \frac{1}{2}\left[\sum_{i,\mu=x,y} J_{i,i+\mu}(\Delta_{\mu} \theta_i)^2 +\sum_{i,\mu=\pm x,\pm y}4J_{i,i+\mu} L_i^2\right]
\ee
where $L_i$ are the quantum operators canonically conjugated to the phase operators $\theta_i$ and $\Delta_{\mu=x,y}$ is the discrete phase gradient in the $\mu$ direction. In the homogeneous case $J_{i,i+\mu}=J$ one can easily see that Eq.\  \pref{quant} describes a sound-like phase mode. Indeed, deriving the 
action corresponding to the Hamiltonian  via the usual identification $ 16 J L_i^2\rightarrow (\hbar \pd_t \theta)^2/16J$ one immediately obtains
\bea
S_{HP}&=&\frac{1}{2}\int dt d\bx  \left[ -J(\nb \theta)^2+\frac{1}{16J\xi_0^2}(\hbar \pd_t \theta)^2\right]=\\
\lb{action}
&=&\frac{1}{2}\int d\omega d\bq \left[ -J\bq^2+\frac{1}{16J\xi_0^2}(\hbar \omega)^2\right]|\theta(\bq,\omega)|^2,
\eea
so that the phase-modes spectrum is sound-like with typical velocity $v_s={4J\xi_0}/\hbar$:
\be
\lb{sound}
\omega=(4J \xi_0/\hbar) q=v_sq.
\ee
Here the scale $\xi_0$ is the coherence length, which represents the typical length scale over which the coarse-grained model \pref{hxy} is valid. 
For Al nanograins the effective quantum phase model is usually taken\cite{ribeiro2014,garcia_prb14} as a Josephson-junction array in the limit where the self-capacitance dominates over the capacitance of the junctions, so that non-local $L_iL_j$ quantum terms in Eq.\ \pref{quant} are absent. In this regime the long-range part of the Coulomb force is screened, and the phase mode is not converted into a plasmon\cite{fazio_review01,halperin_prb89}. For samples in the metallic limit, as the one we are considering, one should also account for the additional dissipation mechanism provided by tunnelling of electrons between neighbouring grains.\cite{halperin_prb89,ribeiro2014}  This effect renormalizes the local capacitance to the effective scale $1/J$,\cite{ribeiro2014,garcia_prb14} exactly as found in Eq.\ \pref{quant} starting directly from the quantum pseudospin model \pref{hxy}.

In weakly-disordered and homogeneous BCS superconductors the phase modes are optically inert at Gaussian level. Indeed, despite the fact that they are crucial to restore the gauge invariance of the longitudinal response\cite{schrieffer,cea_prb14,seibold_cm17}, they do not contribute to the transverse physical response, which reduces to the superfluid contribution at $\omega=0$ . As a consequence the real part of the optical conductivity is well described by the usual Mattis-Bardeen expression\cite{mb}, with a superfluid delta peak at $\omega=0$, followed by a finite absorption above  the threshold $2\Delta$ where Cooper-pair breaking by the electro-magnetic field begins. However, this picture is modified at strong disorder, and in general when the system displays a spatial inhomogeneity, since the disordered phase modes give rise to  finite-frequency absorption\cite{stroud_prb00,cea_prb14,trivedi_prx14} already in the Gaussian approximation. A simple argument to understand this effect is sketched in Fig.\ \ref{fig-phasemode}. In granular Al, shell effects induce a spatial inhomogeneity in the local stiffness of the model \pref{hxy}. In this situation the phase fluctuations of the local SC order parameter induced by the incoming radiation are larger when the phase rigidity (i.e. $J_{ij}$) is smaller. This leads to an inelastic response which results in a partial absorption of the e.m. field, i.e. in a finite value of the real part of the ac optical conductivity.  In other words, the phase modes at the characteristic finite momentum $\xi_0 \bar q\sim 1 $ set by the inhomogeneity become mixed to the transverse physical response at $\bq=0$. The typical energy scale of the absorption is then
\be
\lb{oest}
\hbar \bar \omega \sim 4J (\xi_0 \bar q)\sim 4J 
\ee
so that it is given by the overall scale $J$ of the local Josephson couplings in Eq.\ \pref{hxy}, which will be in the following our main fitting parameter. 

The above qualitative arguments can be made quantitative by computing explicitly the optical conductivity of the model \pref{hxy} within the Holstein-Primakoff approximation \pref{quant}. Following the general procedure outlined in Ref.\ \cite{cea_prb14} the optical conductivity e.g. in the $x$ direction is given by (details are given in Appendix B) 
\bea
\lb{som}
\sigma(\omega)&=&D_s\delta(\omega)+\s_{reg}(\omega),\\
D&=&\frac{4}{N}\sum_i J_{i,i+\hat x}\\
D_s&=&D-\frac{2}{\pi}\sum_\a Z_\a,\\ 
\s_{reg}(\omega)&=&\sum_\a Z_\a \left[\d(\omega-E_\a)+\d(\omega+E_\a)\right]
\eea
with $N$ number of lattice sites and $D$, $D_s$ denote the diamagnetic and superfluid weight, respectively.
Here $E_\a$ are the eigenvalues of the disordered phase spectrum of the model \pref{quant}, and $Z_\a$ represents their electrical dipoles. These are given explicitly by
\be
\lb{za}
Z_\alpha=\frac{4}{E_\alpha}\left(\sum_i J_{i,i+\hat x}\Delta_\mu \phi_{\alpha,i}\right)^2,
\ee
where the $\phi_{\alpha,i}$ are connected to the eigenvectors of each $E_\alpha$ mode (see Eq.\ \pref{phia} below).
When the system is homogeneous, i.e. $J_{i,i+\mu}=J$, each $Z_\alpha$ is proportional to the total gradient of the phase over the system, which vanishes for periodic boundary conditions. In this case, the optical conductivity Eq.\ \pref{som} consists only of the superfluid peak at $\omega = 0$, with $D_s=D=4J$. However disorder in the local Josephson couplings between grains induces a finite electrical dipole for the phase modes, leading to a regular part $\sigma_{reg}$ of the optical conductivity \pref{som}, responsible for the finite-frequency absorption. The exact form of this absorption depends in general on the disorder of the local stiffnesses $J_{i,j}$ used in the model \pref{hxy}. To make contact with the structure of granular Al, we will compute in the next section the distribution of the $J_{ij}$ due to shell effects in an array of nanograins.

\subsection{Calculation of $J_{ij}$ and P($J_{ij}$)}
As was mentioned previously, we model granular aluminum as an array of nano-grains with different size\cite{deutscher_prb80,garcia_prb14}. This implies that, because of shell-effects, each grain has a different superconducting gap and critical temperature, leading in turn to an inhomogeneous distribution of local stiffnesses. 
The strength of the inhomogeneities is controlled by the effective electron-phonon coupling, the grain size distribution and the strength of the coupling among grains. Typically the smaller and more isolated the grain is, the stronger are the inhomogeneities of the sample. 
 In situations in which BCS theory applies, the distribution of the local stiffness $J_{ij}$ of the model \pref{hxy} is given by the  Ambegaokar-Baratoff expression \cite{Ambegaokar1963}:
 \begin{widetext}
\begin{equation}
\lb{jij}
J_{ij} = \frac{\Delta_i \Delta_j}{\beta}\frac{R_Q}{R_N}\sum_{l = -\infty}^{+\infty}\frac{1}{\sqrt{((\frac{(\pi(2l-1))}{\beta})^2+\Delta_i^2)+((\frac{(\pi(2l-1))}{\beta})^2+\Delta_j^2)}}
\end{equation}
\end{widetext}
where $\Delta_i (L)$ is the value of the superconducting order parameter of a grain of size $L$ and $\beta = (k_B T)^{-1}$ with $k_B$ the Boltzmann's constant. 
We compute $\Delta_i (L)$ by solving numerically the BCS gap equation for a spherical grain of size $L$ where we also took 
  into account that the grain is open, namely, electrons can hop from one grain to another. The hopping is more likely as the normal state resistivity becomes smaller. Effectively this hopping smears out the spectral density that weakens shell effects, especially in the deep metallic regime where the electron hops to another grain very quickly. Details of the calculation are given in the Appendix A. 

The distribution of $J_{ij}$ is ultimately determined by the experimental distribution of grain sizes.  From experiments \cite{Deutscher1973,Lerer2014} we know that for the range of resistances of interest the distribution of sizes is close to a log normal distribution with an average diameter $2$ nm and variance of $0.5$ nm. This is the distribution that we will use in the rest of the paper. Despite the broad distribution of sizes, the fact that the typical size is only $2$nm cast doubts on the applicability of BCS in most grains as for $L \sim 2$ nm the mean level spacing is much larger than the superconducting gap, violating thus Anderson's criterion. However we note that this criterion is only applicable for isolated grains where the spectrum is discrete with no imaginary part. The smearing of the spectral density mentioned above has also the effect of bringing back enough spectral weight close to the Fermi energy so that a mean field approach is applicable. This is especially true in the good metallic limit  we are interested in. 

\begin{figure}
	\centering
	\includegraphics[scale=0.3,angle=0]{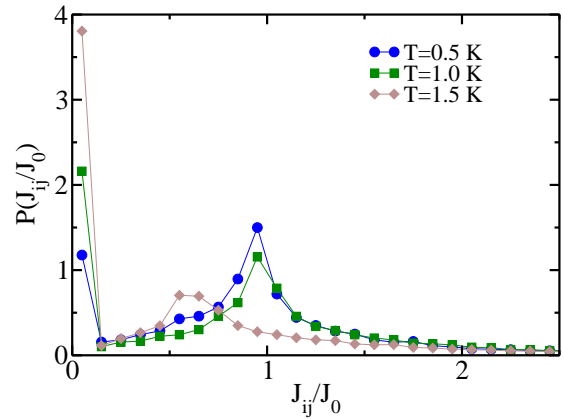}
	\caption{Distribution of the Josephson couplings $P(J_{ij}/J_0)$ for $R_N/R_Q=0.01$, as appropriate for the sample shown in Fig.\ \ref{sig1_exp}. Following the experimental results of Ref.\cite{Deutscher1973,Deutscher1973a,Lerer2014}, we assume that the distribution of grain sizes is log-normal with a typical radius of $1$nm. The couplings $J_{ij}$ are computed exactly  combining the exact solution of the BCS gap equation taking into account the coupling to other grains and the fluctuating spectral density (see Appendix A for more details).}
	\label{fig2}
	\vspace{1cm}
\end{figure}

We are now ready for the computation of the distribution $P(J_{ij})$. We will normalise it with respect to the hopping $J_0$ in the limit of a homogeneous array:
 \be
 \lb{j0array}
 J_{0} = \frac{{\Delta_0}R_Q}{2 R_N} \tanh \left( \frac{\beta {\Delta_0}}{2}\right),
 \ee
where $\Delta_0$ is the bulk value of the gap in the homogeneous case and we defined the quantum of resistance as
\be
R_Q=\frac{h}{4e^2}=6.45 k\O
\ee
while $R_N$ is the normal-state resistance per square of the array. For the sample shown in Fig.\ \ref{sig1_exp} $\rho_{dc}=263\m\O$cm, that for $d=40$ nm gives $R_N=65.75\O\simeq 0.01R_Q$, which is deep in the metallic region. In Fig.~\ref{fig2} we depict the corresponding distribution $P(J_{ij})$ for different temperatures. It is strongly bimodal with a peak at zero hopping corresponding to bonds where the gap vanishes. The other peak has relatively fat tails centred at a $J$ larger than the bulk one at that temperature. Once the distribution of local stiffnesses is known, we can also estimate the global critical temperature as induced by percolation of spheres, which is known to be a good approximation in the limit of small resistance, i.e  $R_N/R_Q \ll 1$ \cite{garcia_prb14}. The existence of large tails in the distribution of the stiffness implies that the global $T_c$ given by percolation is substantially enhanced. Indeed, by mapping the local $J_{ij}$ to local $T_c$ values one can estimate the $T_c$ of the array as the temperature where the percolation threshold is reached, as discussed  in Ref.\ \cite{garcia_prb14}. By using this approach we get here $T_c \sim 1.75 T_c^{\rm bulk}$.  Even if this is still smaller than the experimental value $T_c/T_c^0=2.3$ ($T_c=2.74$ K) it goes in the right direction.


\section{Comparison between theory and experiments}

\begin{figure}
	\centering
	\includegraphics[scale=0.3]{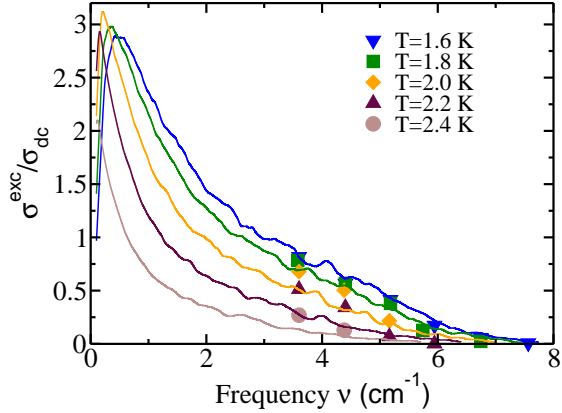}
	\caption{Optical conductivity (solid lines) obtained by using the $P(J/J_0)$ extracted from the shell model, see Fig.\ \ref{fig2}. The data points correspond to the normalized excess conductivity shown as bars in Fig.\ \ref{sig1_exp}. }
	\label{fig-pj}
\end{figure}

After computing the distribution of the local stiffness, we use it to obtain the optical conductivity according to Eq.\ \pref{som} above. As we mentioned previously, we need to fix the overall scale $J$ of the stiffness, since in general $\sigma(\omega)$ will be different from zero in a range of values 
\be
\lb{range}
0<\omega/J<\gamma
\ee
where $\gamma$ depends on the model for the disorder, i.e. on the form of $P(J_{ij})$. The experiments observe a substantial sub-gap absorption in a range of energies such that $\gamma J\lesssim 2\Delta$, i.e. below the threshold for quasiparticle creation. As a consequence we will fix $J$ in order to have our simulation to overlap with the experimental data. We note that our model only describes the absorption by Goldstone modes. Thus, to make contact with the experimentally-determined excess conductivity \pref{sexcess} we must use a proper normalization. Since our model is taken for computational simplification as purely two-dimensional, the conductivity is a multiple of the 2D quantum of absorption  $\sigma_0=e^2/\hbar=0.25 \times 10^{-3} \O$. To translates it in a 3D conductivity we should then divide our 2D conductivity by a transverse length scale $\tilde d$ of the order of the distance between layers.  Notice that, for the same reason, our overall scale $J$ is expected to be quantitatively lower than the stiffness $J_0$ of the whole array, as given by Eq.\ \pref{j0array} above.


To test the effects of the SC inhomogeneity induced by the granular nature of the film we compare in Fig.\ \ref{fig-pj} the optical conductivity obtained from the $P(J_{ij}/J)$ distribution of Fig.\ \ref{fig2} with the experimental data, by using the overall strength of the equivalent 2D homogeneous coupling $J$ as the only free parameter. Even though the distribution of the $P(J_{ij})$ obtained by considering shell effects is itself temperature dependent due to the temperature variation of the SC order parameter $\Delta_i$, we fixed here its shape at $T=0$ and let the system evolve thermally according to the mean-field solution of the model \pref{hxy}. This approach is equivalent to  account for the effects due to phase fluctuations alone, without including also the ones due to thermal suppression of the pairing, which is not present in the phase-only model \pref{hxy}.  Details on the $T$ dependence are given in Appendix B.  As one can see in Fig.\ \ref{fig-pj} the calculations account reasonably well for the 
experimental data, considering that the distribution of the $P(J_{ij}/J)$ has been determined microscopically without any free parameter, except for the overall scale $J$ setting the local 2D stiffness, that we determine as $J=0.4$cm$^{-1}$. The transverse length scale is relatively small, $\tilde d=0.6$ \AA, but still consistent with our 2D limit.

\begin{figure}
	\centering
	\includegraphics[scale=0.3]{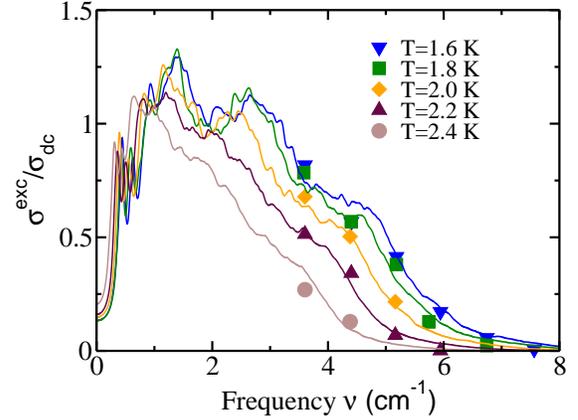}
	\caption{Optical conductivity (solid lines) obtained from the diluted $XY$ model, using a dilution level $p=0.205$ corresponding to a bimodal approximation for the microscopic distribution of Fig.\ \ref{fig2}. The data points correspond to the normalized excess conductivity shown as bars in Fig.\ \ref{sig1_exp}. }
	\label{fig-fit}
\end{figure}

%

By closer inspection of Fig.\ \ref{fig2} one sees that roughly speaking the $P(J_{ij})$ resembles a bimodal distribution, with about $p=0.2$ spectral weight in the bonds with $J_{ij}=0$. We then tested the possibility to reproduce the experiments with a similar but simpler diluted distribution, where $J_{ij}$ has a probability $p=0.2$ of having value $J_{ij}=0$ and probability $1-p$ of having a value $J_{ij}=J$. The result for the bimodal distribution is shown in Fig.\ \ref{fig-fit}.  For the diluted model the support of the $P(J_{ij}/J_0)$ is more compact, so the absorption occurs for a smaller value of $\gamma$ in Eq.\ \pref{range}. This implies that both $J=0.9$ cm$^{-1}$  and $\tilde d=1.24$ \AA \, turn out to be about a factor of two larger. At present, the available experimental data do not allow to seriously discriminate between the two distributions, even though the diluted model displays a slightly better agreement for the temperature evolution. 

Finally, we can use the extracted information on the values of $J$ and $\tilde d$ to make contact with the measured value of the magnetic penetration depth $\lambda$, that can be obtained by extrapolating the measured $\sigma_2(\omega)$ at zero frequency. More specifically, the inverse penetration depth is connected to the superfluid stiffness $J_s=\hbar n_s^{2d}/4m$ of the sample by\cite{pracht_prb16} 
\be
\lb{lambda}
J_s[K]=\frac{0.62\times \tilde d }{\lambda^2[\m m^2]}
\ee
with $\tilde d$ given in\ \AA. According to Eq.\ \pref{som} above, the superfluid stiffness $J_s$ is given by the values of $D_s$ obtained numerically, as $J_s=D_s/4$. For the case of the microscopic $P(J_{ij})$ shown in Fig.\ \ref{fig-pj} we obtain that $J_s=0.625J=0.35$ K, so that 
 $\l^{-2}=0.93$ ($\m$m)$^{-2}$, while for the diluted model we have that $J_s=0.57J=0.72$ K but since $\tilde d$ is larger we get analogously $\l^{-2}=0.9$ ($\m$m)$^{-2}$. In both cases we then obtain a very good agreement with the measured value $\l^{-2}_{exp}=1.06$ ($\m$m)$^{-2}$, reinforcing the consistency of the overall theoretical analysis. 
 
\section{Conclusions}
Our work presents a comprehensive experimental and theoretical analysis of the unconventional superconducting THz response  of granular Al.  By measuring the complex conductivity of the system across the critical temperature we have clearly established the existence of a sizeable optical absorption that cannot be ascribed neither to thermally excited quasiparticles nor to the Cooper pairs broken by the electromagnetic field. Such extra absorption represents a striking violation of the usual Mattis-Bardeen paradigm for disordered BCS superconductors. At the same time, the critical temperature of the film greatly exceeds the one of bulk Al. We have shown that both features can be quantitatively described by modelling granular aluminum as an array of Al nano-grains, where the local Josephson couplings $J_{ij}$ between neighboring grains present  a wide distribution with fat tails. Such inhomogeneity, induced by  shell effects in each nano-grain and by the distribution of grain sizes, can be computed microscopically by using as input parameters the well-known BCS values of bulk Al. The inhomogeneity of the array has two crucial consequences. On one side, the non-vanishing probability of large local values of the SC gap and then of the local Josephson coupling $J_{ij}$ explains, within a percolative scheme, the enhancement of $T_c$ with respect to the homogeneous case. On the other side the inhomogeneity makes the SC phase mode optically active, explaining the anomalous sub-gap absorption. While the SC inhomogeneity has been also proven to emerge in homogeneously disordered films in proximity of the insulating state,  in granular Al film it is a natural consequence of the confinement of superconductivity at the nanoscale. As such, it also manifests in arrays of well-coupled grains, where the overall resistivity still preserves a metallic behavior. The nanostructure has also another advantage, that has been emphasized in the above discussion: it makes the local charging effects stronger, screening the long-range Coulomb forces and leaving intact the sound-like dispersion of the Goldstone mode. The inherent inhomogeneity can then make the phase mode optically active, explaining the anomalous sub-gap absorption observed experimentally. 

Our results provide thus strong evidence that the Goldstone mode can be observed in metallic superconducting nano-grains, provided that the superconducting state is sufficiently inhomogeneous. The optical signatures of the phase modes are not universal, but depend on the probability distribution of the local, inhomogeneous Josephson couplings. On a wider perspective, one can imagine to design an artificial array of Josephson junctions to explore the evolution of the phase response as a function of the Coulomb screening, controlled by the relative strength of the local charging energy with respect to the junction capacitance. This approach would ultimately address the long-standing issue of the interplay between inhomogeneity and Coulomb interactions  in homogeneously disordered systems, and their effect on the optical visibility of the Goldstone mode.

\acknowledgements

This project has been supported by the Deutsche Forschungsgemeinschaft (DFG),
 the German-Israeli-Foundation (GIF), the Italian MIUR (PRINRIDEIRON-2012X3YFZ2), the Italian MAECI under the Italian-India collaborative project SUPERTOP-PGR04879, and the Graphene Flagship. U.S.P. acknowledges a fellowship of the Studienstiftung des deutschen Volkes.  AMG acknowledges partial financial support from a QuantEmX grant from ICAM and the Gordon and Betty Moore Foundation through Grant GBMF5305

%
%
%
%
%
%


\appendix
\section{Numerical calculation of the superconducting gap in isolated and open spherical nano-grains}

We have carried out a fully numerical calculation of the BCS energy gap with an effectively complex spectrum that accounts, in a spherical grain of radius $R$, for the possibility of tunnelling to other grains. 
The resulting gap equation is 
\begin{equation} \label{gapnum}
1=\frac{\lambda_{\rm eff} }{2}\int_{-\epsilon_D}^{\epsilon_D}\sum_n \frac{c_n\nu_\gamma(\epsilon'-\epsilon_n)F(T)}{\sqrt{\epsilon'^2+\Delta(R,T)^2}}d\epsilon'
\end{equation}
with $\lambda_{\rm eff} = \lambda(1+\frac{3\pi}{4k_F R})$. This additional factor accounts for the leading correction to the bulk coupling constant as a consequence of non-trivial matrix elements \cite{Garcia-Garcia2008,Garcia-Garcia2011}, 
\begin{equation}
\nu_\gamma(\epsilon)=\frac{\gamma}{\pi}\frac{1}{\epsilon^2+\gamma^2}.
\end{equation}
The spectrum $\epsilon_n$ is obtained from the zeros of the Bessel function of order $n+1/2$. In the Fermi golden rule approximation, 
\begin{equation}
\gamma \sim  \frac{4zR_Q}{R_N\nu_{TF}(\epsilon_{\rm F})},
\end{equation}
and $F(T) = \tanh (\sqrt{\epsilon'^2+\Delta(R,T)^2}/2T)$.

For the case of granular Al considered in the present work we use the well-known parameters for bulk Al, i.e. $\lambda \approx 0.18$, 
$\epsilon_D =36.4$meV, $k_F \approx 17.5 {\rm nm}^{-1}$ and $\epsilon_F =11.63$ eV, $\xi \sim 1600$nm. We can then compute numerically the superconducting gap $\Delta(R,T)$  as a function of the grain size distribution and of the normal resistance $R_N$, that is controlled by the tunnelling rate. Once determined the local $\Delta_{ij}$ the Josephson coupling between neighboring grains are given by Eq.\ \pref{jij} in the main text.

\section{Calculation of the optical conductivity in the disordered $XY$ model}
Let us first of all show the derivation of Eq.\ \pref{quant} from the quantum pseudospin model \pref{hxy}. As a starting point we compute the mean-field ground state, obtained by assuming that the spins align in the plane along a given, say $x$, direction $\langle S_i^x\rangle\neq 0$. By introducing the Weiss field $B_i=\frac{1}{2}\sum_j J_{ij} \langle S_j^x\rangle$ the Hamiltonian \pref{hxy} can be approximated at mean-field level with $\mathcal{H}_{MF}=-\frac{1}{2}\sum_i B_i S_i^x$. One can then easily derive the self-consistent equations 
\be
\langle S_i^x\rangle=\frac{1}{2}\tanh(\beta B_i)
, \quad B_i=\sum_{j=i+z} J_{ij}\tanh (\beta B_j)
\lb{sground}
\ee
where the sum over $j$ in the above equation is extended to all the $z=4$ nearest-neighbours. Once the ground state has been determined,  fluctuations above it can be described in the Holstein-Primakoff approximation. At $T=0$ this corresponds to introducing the bosonic annihilation (creation) operators $a_i$ ($a^\dagger_i$), related to the spins by:
	\begin{equation}\label{HP_approximation}
	\begin{matrix}
		S^x_i=1/2-a^\dagger_ia_i\quad\quad\quad\quad\quad\quad\quad\\
		-S^z_i+iS_i^y=\left(1-a^\dagger_ia_i\right)^{1/2}a_i\simeq a_i\\
		-S^z_i-iS_i^y=a^\dagger_i\left(1-a^\dagger_ia_i\right)^{1/2}\simeq a^\dagger_i
		\end{matrix}
		\quad,
	\end{equation}
	
where we oriented explicitly the quantization axis along $x$. 
Substituting \eqref{HP_approximation} into \eqref{hxy}, at Gaussian level we get the following Hamiltonian:
\begin{equation}\lb{H_harm}
\mathcal{H}_{XY}\simeq \mathcal{H}_{MF}+\frac{1}{2}\sum_{ij}\left(A_{ij}a^\dagger_ia_j+B_{ij}a_ia_j\right)+h.c.\quad,
\end{equation}
with 
\bea
\lb{aij}
A_{ij}&=&2\delta_{ij}\sum_{j=i+z}J_{ij}-J_{ij}\\
\lb{bij}
B_{ij}&=&J_{ij}
\eea
At finite temperature the form of the Holstein-Primakov relations \eqref{HP_approximation} has to be modified, to remove the contribution of the thermal excitations leading to the suppression of the order parameter \pref{sground} from the definition of the bosonic operators $a$. This implies that Eq.s \pref{HP_approximation} have to be replaced by
	\begin{eqnarray}
		& &S^x_i=\tanh(\beta B_i)/2-a^\dagger_ia_i\nn\\
		& &-S^z_i+iS_i^y=\sqrt{\tanh(\beta B_i)}\left(1-a^\dagger_ia_i\right)^{1/2}a_i\simeq \sqrt{\tanh(\beta B_i)}a_i\nn\\
		& &-S^z_i-iS_i^y=a^\dagger_i\sqrt{\tanh(\beta B_i)}\left(1-a^\dagger_ia_i\right)^{1/2}\simeq \sqrt{\tanh(\beta B_i)}a^\dagger_i\nn\\
				\label{HP_approximationT}
	\end{eqnarray}
At the same time we have to rescale the matrix elements \pref{aij}-\pref{bij} appearing in \eqref{H_harm} as:
\bea
\lb{aijT}
A_{ij}&=&2\delta_{ij}\sum_{j=i+z}J_{ij}-J_{ij}\sqrt{\tanh(\beta B_i)\tanh(\beta B_j)}\\
\lb{bijT}
B_{ij}&=&J_{ij}\sqrt{\tanh(\beta B_i)\tanh(\beta B_j)}
\eea

The Hamiltonian \eqref{H_harm} can be diagonalized via a standard Bogolubov transformation for bosons: 
\bea
\lb{ai}
a_i=\sum_\alpha \left(u_{\alpha i}\gamma_\alpha +v_{\alpha i}\gamma^\dagger_\alpha\right)\\
\lb{ai+}
a^+_i=\sum_\alpha \left(v_{\alpha i}\gamma_\alpha +u_{\alpha i}\gamma^\dagger_\alpha\right)
\eea
so that 
\be
\lb{hdiag}
\mathcal{H}_{PS}=\sum_\alpha E_\alpha\gamma^\dagger_\alpha\gamma_\alpha\text{ ,}
\ee
with the energies $E_\alpha\ge0$ and the coefficients $u$, $v$ determined by solving the secular equations: $\left[\gamma_\alpha,\mathcal{H}_{PS}\right]=E_\alpha \gamma_\alpha$.

To describe the bosonic excitations in terms of collective modes and complete the mapping into the quantum Hamiltonian \pref{quant} we need to make one step further by defining the ''phase operators'' as:
\bea
\lb{PHASE_OP}
\theta_i&\equiv& -\frac{S^y_i}{\tanh(\beta B_i)/2}=\sum_\alpha \frac{\phi_{\alpha i}}{\sqrt{2}i}\left(\gamma^\dagger_\alpha-\gamma_\alpha\right),\\
\lb{phia}
\phi_{\alpha i}&\equiv& \sqrt{\frac{2}{\tanh(\beta B_i)}}\left(	u_{\alpha i}-	v_{\alpha i}\right)
\eea
 The $\theta_i$'s are the quantum operators associated with the phase fluctuations of the SC order parameter: this identification, that will be formally justified below by the coupling of the Gauge field in the original Hamiltonian \eqref{hxy}, can be understood 
 from a semi-classical argument. Since, as we explained below Eq.\ \pref{hxy}, the operator associated with to the local order parameter is $S^-=S^x-iS^y$ if we put $S^-\simeq |\Delta_i|\left(1+i\theta_i\right)$, this obviously implies \eqref{PHASE_OP}, since in the pseudospin mapping $ |\Delta_i|=\langle S^x_i \rangle$. Once  the phase operators $\theta_i$ have been identified, we can equally define their conjugate operators 
 \be
 \lb{lop}
 L_i=-S^z_i=\sum_\alpha \frac{\ell_{\alpha i}}{\sqrt{2}}\left(\gamma^\dagger_\alpha-\gamma_\alpha\right)\quad,
 \ee
 with $\ell_\alpha\equiv (u_{\alpha i}+v_{\alpha i})/\sqrt{2}$,  which satisfy the usual commutation relations $[\theta_i,L_j]=i\delta_{ij}$. By means of the definitions \pref{PHASE_OP}-\pref{lop} one can easily derive Eq.\ \pref{quant}, where the coefficient of the $(\nabla_\mu \theta_i)^2$ term is replaced in general at finite temperature by $J_{i,i+\mu}\ra J_{i,i+\mu}\tanh(\beta B_i)/2\tanh(\beta B_{i+\mu})/2$. 

To derive the current-current response function we need to couple the superconductor to the gauge field $\textbf{A}$. By exploiting once more the mapping between the fermions and the pseudospin operators, and by using the well-known 
Peierls substitution $c_i\ra c_ie^{-ie\int^{\br_i} \bA\cdot d{\bf l}}$ for the fermionic operators, we immediately get that the gauge field enters the pseudospin Hamiltonian \pref{hxy} as
\begin{equation}\label{Peierls}
S^+_iS^-_{i+\mu}\rightarrow S^+_iS^-_{i+\mu}e^{-2ieA^\mu_i}\quad,
\end{equation}
 with the factor of two taking into account the double charge of a Cooper-pair and $A_i^\mu\equiv \bA(\br_i)\cdot \hat \br_\mu$. Here $e$ is the electron charge, and we set the sound velocity  $c=1$. By means of the relation \pref{Peierls} one can derive the current operator in terms of the pseudospin operators, and then use again the Holstein-Primakoff transformations \pref{HP_approximationT} to express it in terms of the spin-wave bosonic operators $a_i,a_i^\dagger$, or equivalently the phase-momentum operators \pref{PHASE_OP}-\pref{lop}. In particular, it can be easily shown that \eqref{Peierls} is equivalent to substitute $\Delta_\mu\theta_i\rightarrow\Delta_\mu\theta_i-2eA^\mu_i$ inside the Hamiltonian \eqref{quant}, thus justifying a posteriori the role of the $\theta_i$ in representing the phase of the SC order parameter.

The operator associated with the current flowing across the link $(i,i+\mu)$ then reads:
 \begin{equation}
 	I^\mu_i=-\frac{\partial \mathcal{H}_{PS}}{\partial A^\mu_i}=-2eJ^\mu_i\left(2eA^\mu_i-\Delta_\mu\theta_i\right)\quad.
 \end{equation}
 where, as usual, the first term is the diamagnetic part, linearly proportional to the gauge field, while the second one defines the paramagnetic current operator. By using the Kubo formula, the current in linear-response theory can then be written as 
\begin{equation}\label{linear_R}
 \langle I^\mu_i(\omega)  \rangle=-\sum_{j\nu} K^{\mu\nu}_{ij}(\omega)A^\nu_j(\omega),
\end{equation}
where the electromagnetic kernel $K^{\mu\nu}_{ij}$ is computed explicitly as:
 \begin{equation}
    K^{\mu\nu}_{ij}(\omega)=4e^2\left[ J^\mu_i\delta^{\mu\nu}\delta_{ij}- J^\mu_i J^\nu_j\sum_\alpha\frac{\Delta_\mu\phi_{\alpha i}\Delta_\nu\phi_{\alpha j}E_\alpha}{E_\alpha^2-(\omega+i0^+)^2}			\right].
 \end{equation}

In the case of a uniform field along the $x$ axis $\mathbf{A}=A(t)\hat{x}$, we can define the optical conductivity by averaging over the space coordinates: $\sigma(\omega)=-\frac{i/N}{(\omega+i0^+)}\sum_i\frac{\langle I^x_i(\omega)  \rangle}{ A(\omega)}$, with $N$ the number of lattice sites. The real part of the conductivity reduces then to Eq.\ \pref{som} in the main text. To compute it we determine numerically at each temperature the eigenvalues $E_\alpha$ and the eigenvectors \pref{ai}-\pref{ai+} of the Hamiltonian \pref{hdiag} for a given disorder configuration of the $J_{ij}$ couplings (using $N=30$), and we then average the result over 100 disorder configurations.
The self-consistent equation \pref{sground} gives a vanishing order parameter for a given mean-field temperature $T_{MF}$. 
To make a direct comparison with the experimental data we then plot our numerical results as a function of $T/T_{MF}$ for the corresponding value of $T/T_c$ of the experimental measurements.

\end{document}